\documentclass[twocolumn]{aastex7}

\usepackage{amsmath}

\newcommand{\kms}{km~s$^{-1}$}

\newcommand{\HeI}{He~{\sc i}}

\newcommand{\CII}{C~{\sc ii}}
\newcommand{\CI}{C~{\sc i}}
\newcommand{\NaI}{Na~{\sc i}}

\newcommand{\MgII}{Mg~{\sc ii}}

\newcommand{\SiII}{Si~{\sc ii}}
\newcommand{\SiIII}{Si~{\sc iii}}

\newcommand{\SII}{S~{\sc ii}}
\newcommand{\CaII}{Ca~{\sc ii}}

\newcommand{\FeII}{Fe~{\sc ii}}
\newcommand{\FeIII}{Fe~{\sc iii}}
\newcommand{\CoII}{Co~{\sc ii}}

\newcommand{\NiII}{Ni~{\sc ii}}

\newcommand{\peakMJD}[0]{60525.5}
\begin{document}

\title{Early and Extensive Ultraviolet Through Near Infrared Observations of the Intermediate-Luminosity Type Iax Supernovae 2024pxl}

\correspondingauthor{W.~B.~Hoogendam}

\author[0000-0003-3953-9532]{W.~B.~Hoogendam}
\altaffiliation{NSF Graduate Research Fellow}
\affiliation{Institute for Astronomy, University of Hawai'i at Manoa, 2680 Woodlawn Dr., Hawai'i, HI 96822, USA }
\email[show]{willemh@hawaii.edu} 

\author[0000-0002-5221-7557]{C.~Ashall}
\affiliation{Institute for Astronomy, University of Hawai'i at Manoa, 2680 Woodlawn Dr., Hawai'i, HI 96822, USA }
\email{cashall@hawaii.edu}

\author[0000-0002-6230-0151]{D.~O.~Jones}
\affiliation{Institute for Astronomy, University of Hawai'i, 640 N. A'ohoku Pl., Hilo, HI 96720, USA}
\email{dojones@hawaii.edu}

\author[0000-0003-4631-1149]{B.~J.~Shappee}
\affiliation{Institute for Astronomy, University of Hawai'i at Manoa, 2680 Woodlawn Dr., Hawai'i, HI 96822, USA }
\email{shappee@hawaii.edu}

\author[0000-0002-2471-8442]{M.~A.~Tucker}
\altaffiliation{CCAPP Fellow}
\affiliation{Center for Cosmology and Astroparticle Physics, The Ohio State University, Columbus, OH, USA}
\affiliation{Department of Astronomy, The Ohio State University, Columbus, OH, USA}
\email{tucker.957@osu.edu}

\author[0000-0003-1059-9603]{M.~E.~Huber}
\affiliation{Institute for Astronomy, University of Hawai'i at Manoa, 2680 Woodlawn Dr., Hawai'i, HI 96822, USA }
\email{mehuber7@hawaii.edu}

\author[0000-0002-4449-9152]{K.~Auchettl}
\affiliation{School of Physics, The University of Melbourne, Parkville, VIC, Australia}
\affiliation{Department of Astronomy and Astrophysics, University of California, Santa Cruz, CA, USA}
\email{katie.auchettl@unimelb.edu.au}

\author[0000-0002-2164-859X]{D.~D.~Desai}
\affiliation{Institute for Astronomy, University of Hawai'i at Manoa, 2680 Woodlawn Dr., Hawai'i, HI 96822, USA }
\email{dddesai@hawaii.edu}

\author[0000-0003-3429-7845]{A.~Do}
\affiliation{Institute of Astronomy, Cambridge, UK}
\affiliation{Kavli Institute for Cosmology, Cambridge, UK}
\email{ajmd6@cam.ac.uk}

\author[0000-0001-9668-2920]{J.~T.~Hinkle}
\altaffiliation{FINESST FI}
\affiliation{Institute for Astronomy, University of Hawai'i at Manoa, 2680 Woodlawn Dr., Hawai'i, HI 96822, USA }
\email{jhinkle6@hawaii.edu}

\author[0009-0005-5121-2884]{M.~Y.~Kong}
\affiliation{Institute for Astronomy, University of Hawai'i at Manoa, 2680 Woodlawn Dr., Hawai'i, HI 96822, USA }
\email{ykong2@hawaii.edu}

\author[0009-0003-8153-9576]{S.~Romagnoli}
\affiliation{School of Physics, The University of Melbourne, Parkville, VIC, Australia}
\email{romagnolis@student.unimelb.edu.au}

\author[0009-0008-3724-1824]{J.~Shi}
\affiliation{School of Physics, The University of Melbourne, Parkville, VIC, Australia}
\email{jennifer.shi@student.unimelb.edu.au}

\author[0009-0000-6821-9285]{A.~Syncatto}
\affiliation{Department of Physics and Astronomy, University of Hawai`i at Hilo, Hilo, HI, USA}
\email{jaq45@hawaii.edu}

\author[0000-0002-5740-7747]{C.~D.~Kilpatrick}
\affiliation{Center for Interdisciplinary Exploration and Research in Astrophysics, Northwestern University, Evanston, IL, USA}
\email{ckilpatrick@northwestern.edu}

\begin{abstract}

We present ultraviolet (UV) through near-infrared (NIR) photometric and spectroscopic observations of the nearby SN~2024pxl, the third Type Ia supernova (SN~Ia) in NGC~6384. 
SN~2024pxl is a Type Iax supernova (SN~Iax) with an intermediate luminosity ($M_r = -16.99\pm0.32$~mag) and an average SN~Iax light curve decline rate. % ($\Delta m_{15}(r)=0.69\pm0.05$~mag)
SN~2024pxl was discovered $\sim$3~days after first light, and the rising light curve follows a single power law that is inconsistent with significant interaction with a companion star or circumstellar material.
Our extensive NIR photometric coverage is comparable to that of the well-observed SNe~Iax 2005hk and 2012Z, and we demonstrate that the $J-H$ colors of SNe~Iax differ from normal SNe~Ia and appear to be more homogeneous as a class.  
Spectroscopically, we report the earliest-ever NIR spectrum of a SN~Iax as measured from maximum light ($t\approx-9$~days): a featureless continuum with similarities to a $\sim$9\,000~K blackbody, and the line velocities are consistent with a mixed-ejecta structure, with C, Si, and Fe having similar velocities and velocity evolutions. 
We find a tentative correlation between the $H$-band break \CoII\ velocity $\sim$20~days post-peak and absolute magnitude, with more luminous SNe~Iax showing faster \CoII\ velocities.
Our observations suggest that SN~2024pxl resulted from the thermonuclear disruption of a CO white dwarf star that undergoes deflagration burning. 

\end{abstract}


\section{Introduction} \label{sec:introduction}

Type Ia supernovae (SNe~Ia) are a large class of transients resulting from the explosion of carbon-oxygen white dwarf (CO-WD) stars \citep{hoy60}. A non-negligible fraction (5-30\% of the SN~Ia rate; \citealp{Li2011, White15, Srivastav22, Desai24}) of these events are the peculiar Type Iax supernovae (SNe~Iax; \citealp[see][]{Foley13, Jha17}), with SN~2002cx \citep{li03} serving as the prototypical member of the class.\footnote{SNe~Iax have an eponymous alternative name of ``2002cx-like SNe~Ia.''} These explosions are generally less luminous and show lower ejecta velocities than normal SNe~Ia. 

A favored progenitor scenario for SNe~Iax is the single-degenerate scenario with one WD and one non-degenerate companion \citep{Whelan73}. Unlike for normal SNe~Ia, this scenario has robust observational support based on the progenitor detection of SN~2012Z \citep[][]{McCully14b}. Furthermore, two SNe~Iax, SNe~2004cs and 2007J, show strong \HeI\ emission at $\sim$+40~days after maximum light \citep[something not seen in normal SNe~Ia \citealp{Tucker20}][although see also \citealp{White15} and \citealp{Foley16_Iax_Late_Time}]{Foley13}, which has been argued to support a non-degenerate He-star companion \citep{Foley13}. 
SNe~Iax are predominantly found in late-type host galaxies \citep{Lyman18}, whereas normal SNe~Ia occur in both late- and early-type galaxies \citep{Galbany22}. Late-type galaxies may preferentially form proposed progenitor systems necessary for SNe~Iax, such as binary systems composed of a CO WD accreting from a He star companion \citep{Piersanti14}. These systems are favored in part due to pre- and post-explosion \emph{Hubble Space Telescope (HST)} imaging of SN~2012Z that reveals a blue progenitor \citep{McCully14b} and a late-time flux excess which may be consistent with a bound remnant, late-time circumstellar medium interaction, or a peculiar late-time decay \citep{McCully22, Schwab25}.

Many studies favor the pure deflagration of a CO WD to explain SNe~Iax \citep[e.g.,][]{Branch04, Phillips07, Sahu08, Jordan12b, Kromer13a, Fink14, Bulla20, Leung20, Zeng20, Magee22, Lach22a, Feldman23, Maguire23}, where the flame front propagates as a deflagration wave. This deflagration front is subsonic and produces large-scale instabilities that mix elemental distributions within the ejecta \citep[e.g.,][]{Gamezo04}. 
Alternatively, pulsational delayed detonation models \citep[e.g.,][]{Ivanova74, Hoeflich95, Baron12, Stritzinger15} have been invoked to explain brighter SNe~Iax, in particular SN~2012Z. In this explosion mechanism, the CO WD pulsates once before thermonuclear runaway. The flame front stars as a deflagration and then transitions into a supersonic detonation. The primary difference between these two models lies in the ejecta stratification, with deflagration explosions producing fully-mixed ejecta and pulsational delayed detonations having a layered outermost ejecta. 
 
Photometrically, SNe~Iax are bluer in the ultraviolet than normal SNe~Ia before maximum brightness and then redder than normal SNe~Ia $\sim$10 days after maximum brightness \citep{Milne10}. They lack the secondary peaks in their $r$-, $i$-, and near-infrared (NIR) light curves commonly seen in normal SNe~Ia \citep{Phillips07, Gonzalez14, Ashall20}.
Spectroscopically, the ejecta velocities are slower than normal SNe~Ia and range from 2\,000 km s$^{-1}$ to 7\,000 km s$^{-1}  $ \citep[e.g.,][compared to $\sim$10\,000 \kms\ at peak light for normal SNe~Ia; \citealp{Morrell24}]{Stritzinger09_CBET, Foley13, Jha17}, and signatures of C are common in SNe~Iax, indicating some primordial CO WD material remains unburnt \citep{Foley13}. 

SNe~Iax are generally featureless in their early NIR spectra \citep[e.g.,][]{Stritzinger15, Maguire23}, unlike normal SNe~Ia \citep[e.g.,][]{Hsiao13, Hsiao15, Hsiao19, Hoogendam25a} and as early as $\sim$18~days after maximum light the slow ejecta velocities enable the clear resolution of $H$-band \CoII\ features (e.g., SNe 2010ae, 2012Z, and 2014ck \citealp{Stritzinger14, Stritzinger15, Tomasella16}) that are heavily blended in normal SNe~Ia. Comprehensive NIR observations of SNe~Iax are rare; only SN~2012Z has NIR spectral coverage spanning a week before until a month after maximum light. Unlike normal SNe~Ia, which become ``nebular'' with spectra showing solely forbidden lines, SNe~Iax never reach this stage, and even a year past maximum show permitted \FeII, \NaI~D and \CaII~NIR~triplet lines, indicating a dense, slowly expanding emission region long after the SN explosion which can be naturally explained if some material stays gravitationally bound to the remnant WD \citep{jha06, Sahu08, Foley10a, Foley16_Iax_Late_Time, Camacho-Neves23}. 

In this manuscript, we present extensive UV through NIR observations of the nearby SN~Iax~2024pxl from shortly after discovery (the earliest optical spectrum is at $-10.1$~days and the earliest NIR spectrum at $-9.1$~days, the earliest ever for a SN~Iax) to a month and a half after peak light when it entered a seasonal break. Recent preprints by \citet{Singh25} and \citet{Kwok25} present complementary data of SN~2024pxl to those in our work. \S \ref{sec:data} presents our observations and data reductions, \S \ref{sec:photometry} and \S \ref{sec:spectroscopy} present our photometric and spectroscopic analyses, respectively. \S \ref{sec:conclusion} provides a summary and concluding remarks. Throughout this work, we adopt $H_0 = 72$ km s$^{-1}$ Mpc$^{-1}$, $\Omega_m = 0.30$, and $\Omega_{vac} = 0.70$ for this work. 

\section{Data} \label{sec:data}

\begin{table}[]
    \centering
    \caption{Log of the Photometric Observations of SN~2024pxl. The full table is available online.}\label{tab:phot-data}
    \begin{tabular}{ccccc}
        \hline 
        MJD         & Filter    & $m$       & $\sigma_m$    & Source    \\
        \hline \hline
        60524.32    & $Y$       & 15.36     & 0.01          & UKIRT     \\
        60561.27    & $Y$       & 15.56     & 0.01          & UKIRT     \\
        \vdots      & \vdots    & \vdots    & \vdots        & \vdots    \\
        60608.21    & $Y$       &  16.89    & 0.01          & UKIRT     \\
        \hline \\
    \end{tabular}
\end{table}

\begin{table}[]
    \centering
    \caption{Log of optical and NIR spectroscopic observations. The reported phases are with respect to the time of $g$-band maximum on MJD \peakMJD.}\label{tab:spec-log}
    \begin{tabular}{ccccc}
        \hline 
        UT Date & MJD & Phase [days] &  Telescope & Spectrograph \\ 
        \hline \hline 
                       &               & Optical    &              &           \\
            2024-07-23 &	60515.4	   & $-10.1$    &  ANU~2.3m    &   WiFeS   \\
            2024-07-26 &	60517.3	   & $ -8.2$    &  UH~2.2m     &   SNIFS   \\
            2024-07-28 &	60519.4	   & $ -6.1$    &  UH~2.2m     &   SNIFS   \\
            2024-07-29 &	60520.4	   & $ -5.1$    &  UH~2.2m     &   SNIFS   \\
            2024-07-30 &	60521.4	   & $ -4.1$    &  UH~2.2m     &   SNIFS   \\
            2024-07-31 &	60522.4	   & $ -3.1$    &  UH~2.2m     &   SNIFS   \\
            2024-08-01 &	60523.3	   & $ -2.2$    &  UH~2.2m     &   SNIFS   \\
            2024-08-08 &	60530.3	   & $ +4.8$    &  UH~2.2m     &   SNIFS   \\
            2024-08-09 &	60531.3	   & $ +5.8$    &  UH~2.2m     &   SNIFS   \\
            2024-08-12 &	60534.3	   & $ +8.8$    &  UH~2.2m     &   SNIFS   \\
            2024-08-13 &	60535.3	   & $ +9.8$    &  UH~2.2m     &   SNIFS   \\
            2024-08-16 &	60538.3	   & $ 12.8$    &  UH~2.2m     &   SNIFS   \\
            2024-08-18 &	60540.3	   & $+14.8$    &  UH~2.2m     &   SNIFS   \\
            2024-08-21 &	60543.3	   & $+17.8$    &  UH~2.2m     &   SNIFS   \\
            2024-08-22 &	60544.3	   & $+18.8$    &  UH~2.2m     &   SNIFS   \\
            \hline
                       &                  & NIR            &           &           \\
            2024-07-25 &	60516.4	      & $ -9.1   $     &  Gemini-N &   GNIRS   \\
            2024-08-01 & 	60523.3	      & $ -2.2   $     &  IRTF     &   SpeX    \\
            2024-08-01 &	60523.4	      & $ -2.1   $     &  Gemini-N &   GNIRS   \\
            2024-08-14 &	60536.4	      & $+10.9   $     &  IRTF     &   SpeX    \\
            2024-09-05 &	60558.3	      & $+32.8   $     &  IRTF     &   SpeX    \\
            2024-09-10 &	60563.2	      & $+37.7   $     &  IRTF     &   SpeX    \\
            2024-09-11 &	60564.1	      & $+38.6   $     &  Gemini-N &   GNIRS   \\
            2024-09-14 &	60567.2	      & $+41.7   $     &  IRTF     &   SpeX    \\
            2024-09-15 &	60568.2	      & $+42.7   $     &  Gemini-N &   GNIRS   \\
            2024-09-18 &	60571.2	      & $+45.7   $     &  IRTF     &   SpeX    \\
            \hline \\
    \end{tabular}
\end{table}

\begin{figure*}
    \centering
    \includegraphics[width=\textwidth]{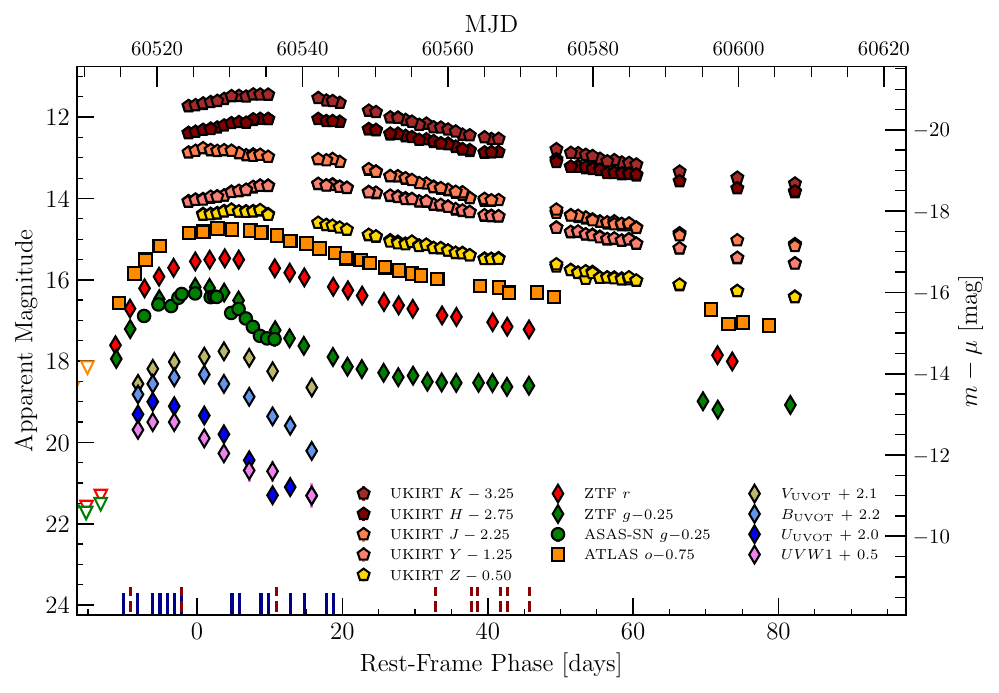}
    \caption{
        The UV, optical, and NIR photometry of SN~2024pxl corrected for Milky-Way extinction. The epochs with optical (NIR) spectra are denoted with solid blue (dotted red) lines. Upper limits are denoted with hollow triangles. The phase is calculated from the time of $g$-band maximum light. 
        }
    \label{fig:photometry-lightcurves}
\end{figure*}

The Zwicky Transient Facility (ZTF; \citealp{Bellm19}) discovered SN~2024pxl on UT 2024-07-23 at 06:34:48 (MJD=60514.3) with a $r$-band magnitude of 17.7 and a last non-detection of $r>19.4$~mag two days prior on UT 2024-07-21 06:04:13 (MJD 60512.3; \citealp{2024pxl_Discovery}). It was classified on UT 2024-07-25 20:10:13 ($\sim$4.5~days later) as a SN~Iax based on a spectrum from the Wide Field Spectrograph (WiFeS; \citealp{Dopita07, Dopita10}) on the Australian National University (ANU)~2.3m telescope \citep{2024pxl_Classification}.

\subsection{Photometric Data}\label{sec:phot-data}

We collected optical survey photometry of SN~2024pxl from the ZTF, the Asteroid Terrestrial-impact Last Alert System \citep[ATLAS;][]{Tonry18}, and All-Sky Automated Search for SuperNovae \citep[ASAS-SN][]{Shappee14, Kochanek17, Hart23} surveys. We obtained the ZTF data using the ZTF forced photometry service \citep{Masci23}.\footnote{\href{https://irsa.ipac.caltech.edu/Missions/ztf.html.}{https://irsa.ipac.caltech.edu/Missions/ztf.html.}} We retrieved the ATLAS data from the  ATLAS Transient Science Server \citep{smith20}. ATLAS data taken on the same night were stacked using a weighted average. ATLAS only observed SN~2024pxl using the $o$ band. Finally, we obtained ASAS-SN photometry from the SkyPatrol service \citep{Kochanek17, Hart23}.\footnote{Accessed via \href{https://asas-sn.osu.edu}{https://asas-sn.osu.edu}.}

We also obtained \emph{ZY\!JHK} photometry with the UKIRT 3.8m telescope using the Wide Field Camera (WFCAM; \citealt{Casali07, Hodgkin09}). We processed these data with {\tt Photpipe} \citep{Rest05} and calibrated them using The Two Micron All Sky Survey (2MASS; \citealp{Skrutskie06}) following \citet{Hodgkin09} and \citet{Peterson23}. We used unforced, non-difference-imaged photometry from \texttt{DOPHOT} \citep{Schechter93}. The background light is negligible compared to the flux from SN~2024pxl at the phases of our photometry, so the lack of a template does not significantly affect our photometry. 

Finally, we obtained UV photometry from the Neil Gehrels Swift Observatory (\emph{Swift}; \citealp{Gehrels04}) using \emph{Swift}'s Ultraviolet/Optical Telescope (UVOT; \citealp{Roming05}). We used \texttt{HEASOFT} version 6.31.1 to combine the images with the \texttt{UVOTIMSUM} package and to obtain aperture photometry with the \texttt{UVOTSOURCE} package. An aperture of 5\arcsec\ is used for the source, and a 15\arcsec\ region centered on $(\alpha,\delta)~=~(17^{\mathrm{h}}32^{\mathrm{m}}28\fs47, +07\degr05\arcmin27\farcs68)$ is used for the background. We report AB magnitude system using the \citet{Breeveld11} zero points. 
The $UVW2$ and $UVW1$ filters have known red leaks in their transmission functions, broadening the photon distribution to include optical photons \citep{Brown10}. 
The host-galaxy light is negligible compared to the SN light for the reddest four \emph{Swift} bands, but there is a non-negligible host background for the $UVW2$ and $UVM2$ bands. After subtracting the median flux from the pre-SN \emph{Swift} images, there are no $5\sigma$ detections in either band. Therefore, we exclude these bands from our analysis. 
Table \ref{tab:phot-data} presents a log of our photometric observations.

\subsection{Spectroscopic Data}

Fourteen optical spectra were taken on the University of Hawaii (UH)~2.2m telescope using the SuperNova Integral Field Spectrograph (SNIFS; \citealt{Lantz04}) as part of the Spectroscopic Classification of Astrophysical Transients (SCAT; \citealt{Tucker22c}) Survey. The spectra were reduced with the SCAT pipeline \citep{Tucker22c}. We also include the classification spectrum, which we re-reduced with \texttt{pyWiFeS} \citep{Childress14}.  Table \ref{tab:spec-log} compiles the optical spectroscopic observations from $-10.1$ to $+45.7$~days from $g$-band maximum.

Of the total ten NIR spectra in our dataset, four were taken on the Gemini North 8.1m telescope using the Gemini Near-InfraRed Spectrograph (GNIRS; \citealt{Elias06a, Elias06b}) and six on the 3m NASA InfraRed Telescope Facility (IRTF) with SpeX \citep{Rayner03}. The GNIRS and SpeX data were reduced with telluric corrections from an A0V star using the standard \texttt{PypeIt} \citep{Prochaska20} and \texttt{Spextool} \citep{Cushing04} procedures, respectively. Table \ref{tab:spec-log} catalogs our NIR spectroscopic observations, which include the earliest ever NIR spectrum of a SN~Iax at $\sim$$-9$~days before $g$-band peak. 

\subsection{Host Galaxy}
SN~2024pxl exploded in NGC~6384, a spiral galaxy at $z=0.005554$ \citep{Haynes98}. NGC~6384 has a Milky Way line-of-sight extinction of $E(B-V)=0.106$~mag \citep{Schlafly11}. We use the \texttt{Blast} tool \citep{Jones24}\footnote{\url{https://blast.scimma.org/}.} to estimate the properties of NGC~6384 using archival host-galaxy photometry from the Galaxy Evolution Explorer \citep{Martin05}, 2MASS \citep{Skrutskie06}, Wide-field Infrared Survey Explorer \citep{Wright10}, Sloan Digital Sky Survey \citep{Fukugita96, York00}, and  Panoramic Survey Telescope and Rapid Response
System \citep{Chambers16} using PSF-matched elliptical apertures. 
% The archival photometry is logged in Table \ref{tab:host-phot}. 
Host-galaxy spectral energy distribution (SED) parameters are estimated using \verb|Prospector| \citep{Leja19, Johnson21, Wang23_prospector}.
For NGC~6384, we find the following global parameters:
a mass of $\log_{10}(M/M_\odot)=10.57^{-0.09}_{+0.06}$,
a star-formation rate (SFR) of $\log_{10}\left(\mathrm{SFR/yr/M_\odot}\right) = -0.69_{-0.48}^{+0.42}$,
a specific star-formation rate (sSFR) of $\log_{10}\left(\mathrm{sSFR/yr}\right) = -11.25_{-0.52}^{+0.44}$, and a mass-weighted mean stellar age of $12.46_{-3.57}^{+0.33}$~Gyr. The galaxy age may be dominated by the bulge light, as this age is rather old for a star-forming galaxy like NGC~6384, and the local age in the area of SN~2024pxl is $\sim$2~Gyr younger than the global age, suggesting that the older bulge likely skews the global age. 

% \begin{table}[]
%     \centering
%     \caption{Log of Archival Host-Galaxy Photometry.} \label{tab:host-phot}
%     \begin{tabular}{ccccc}
%         \hline \\
%         Filter & $F_\lambda$ [mJy] & $\sigma_{F_\lambda}$ &  Mag & $\sigma_{\rm Mag}$ \\ 
%         \hline \hline \\
%             $GALEX_{\mathrm{FUV}}$	 &   112.78      &	6.43      &	18.769    &	0.062 \\
%             $GALEX_{\mathrm{NUV}}$	 &   275.69      &	8.83      &	17.799    &	0.035 \\
%             SDSS $u$	             &   3552.72     &	33.9      &	15.024    &	0.010 \\
%             PanSTARRS $g$	         &   29615.27    &	23.01     &	12.721    &	0.001 \\
%             SDSS $i$	             &   104747.15   &	108.44    &	11.350    &	0.001 \\
%             PanSTARRS $z$	         &   133359.65   &	50.27     &	11.087    &	0.000 \\
%             PanSTARRS $y$	         &   166687.63   &	84.83     &	10.845    &	0.001 \\
%             2MASS $J$	             &   196870.22   &	1076.52   &	9.770 &	0.006 \\
%             2MASS $H$                &   236602.67   &	1633.17   &	9.097 &	0.007 \\
%             2MASS $K$                &   192118.68   &	1630.69   &	8.853 &	0.009 \\
%             $WISE$ $W1$              &   107495.96   &	2111.06   &	8.623 &	0.021 \\
%             $WISE$ $W4$	             &   68447.72    &	7201.95   &	5.192 &	0.114 \\
%         \hline
%     \end{tabular}
% \end{table}

\begin{figure}
    \centering
    \includegraphics[width=\linewidth]{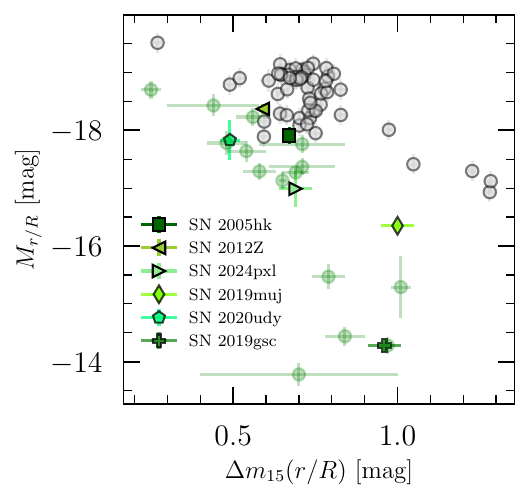}
    \caption{
        $r/R$-band absolute magnitude versus $\Delta m_{15}(r/R)$ for literature SNe~Iax (green circles, with labeled points representing SNe~Iax from \citealp{Foley13, Stritzinger14, Magee16, Tomasella16, Barna21, Karambelkar21, Singh23}) and normal SNe~Ia (grey circles) from the Carnegie Supernova Project \citep{Krisciunas17}. SN~2024pxl is denoted with a light green square and is an intermediate-luminosity SN~Iax.
        }
    \label{fig:photometry-lwr_Iax}
\end{figure}

\subsubsection{SN 2024pxl's Siblings: SNe 1971L and 2017drh}
In this section, we use two previous SNe~Ia, 1971L and 2017drh, to estimate the distance to NGC 6384. A growing number of SN~Ia share a host galaxy with legacy SNe~Ia and are referred to as ``sibling" SNe~Ia (e.g., SNe 1997cn and 2015bo; \citealp{Turatto98, Hoogendam22}; see \citealp{burns20, Kelsey24} for compilations of other SNe~Ia siblings).  

SN~1971L was discovered by \citet{Logan71} on 1971-06-24 with subsequent IAU Circulars from \citet{Locher71} and \citet{Seiler71}, and photometric observations from \citet{Barbon73} and \citet{Lyutyj76}. We omit reporting fit results for SN~1971L; the observations were taken from multiple sites using photometric plates, and there is a large, systematic scatter in the plate photometry that makes SN~1971L unreliable for a precise distance estimate.

SN~2017drh was discovered by \citet{2017drh_discovery} on 2017-05-03. We perform photometry on public images of SN~2017drh from the Las Cumbres Observatory Global Telescope (LCOGT) with \texttt{photpipe} on the pixel-level calibrated LCOGT data. We create difference images for \texttt{photpipe} using the latest observed epoch without SN~Ia flux as the template image. We use SuperNovae in object-oriented Python (\texttt{SNooPy}; \citealp{Burns11, Burns14}) to fit SN~2017drh (with \texttt{EBV\_model2}) to estimate its properties.

For SN~2017drh, we find
a distance modulus of $\mu=32.31\pm0.32$~mag,
light curve shape $s_{BV}=0.89\pm0.03$~mag, 
and host-galaxy extinction of $E(B-V)_{\mathrm{host}}=1.19\pm0.06$~mag.
This distance is consistent with the Tully-Fisher distance of $\mu=32.00\pm0.45$ \citep{Tully16}. We adopt the distance from SN~2017dlh ($\mu=32.31\pm0.32$~mag; $d=28.95\pm5.14$~Mpc) for this work.

\subsubsection{Host-Galaxy Extinction}

In addition to the Milky Way extinction, we estimate the host-galaxy extinction from the \NaI~D feature using the relationship between $E(B-V)_{\rm host}$ from \citet{Poznanski12} with uncertainties calculated using the method from \citet{Phillips13}. We find $E(B-V)_{\rm host} = 0.19\pm0.03$~mag, which we adopt for the rest of this work. Assuming $R_V=3.1$ and using the \citet{Cardelli89} extinction law, this corresponds to $A_g = 0.36$~mag and $A_r = 0.25$~mag.

\section{Photometric Analysis} \label{sec:photometry}

\subsection{Light Curves}\label{sec:lightcurves}
Figure \ref{fig:photometry-lightcurves} shows our photometric observations. We highlight our extensive NIR photometric coverage: SN~2024pxl is the third SN~Iax with a NIR light curve with high cadence over a 3-month timescale after SNe~2005hk \citep{Phillips07} and 2012Z \citep{Yamanaka15}. We find a $g$-band peak of $m_g=16.05\pm0.01$~mag on MJD $60525.5\pm0.2$ and an $r$-band peak of $m_r=15.57\pm0.01$~mag on MJD $60529.2\pm0.2$ using a Gaussian Process fit as implemented in \texttt{SNooPy}.

\citet{Singh23} propose that SN~Iax form two groups based on their absolute magnitude and decline rates \citep[see also][]{Magee16}, with the groups consisting of more luminous SNe~Iax ($-17 \gtrsim M_r \gtrsim -19$~mag) and low-luminosity SNe~Iax ($M_r \gtrsim-15.5$~mag). Figure \ref{fig:photometry-lwr_Iax} presents the $\Delta m_r(15)$ vs $M_{r/R}$ phase space with normal SNe~Ia from the Carnegie Supernova Project \citep{Krisciunas17} and literature SNe~Iax. SN~2024pxl is on the lower edge of the luminous group with $\Delta m_r(15) = 0.69\pm0.05$~mag and $M_r=-16.99\pm0.32$~mag.

\begin{figure*}
    \centering
    \includegraphics[width=\textwidth]{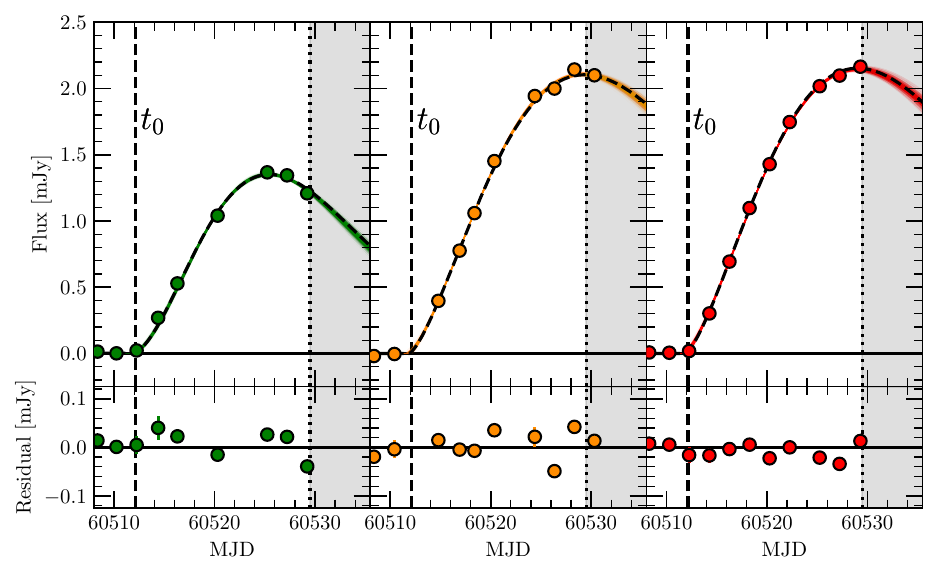} \\
    \caption{
        The rising light curve fits for SN~2024pxl. The dashed black line denotes the time of first light ($t_0$) from the best-fit parameters. The shaded grey region after the dotted black line denotes where data are excluded from the fit. The solid black line represents the best-fit light curve, with the colored lines showing 1000 randomly sampled light curves from our posterior. 
        \emph{Left:}   The ATLAS $g$-band rising light curve.
        \emph{Center:} The ZTF $o$-band rising light curve.
        \emph{Right:}  The ZTF $r$-band rising light curve.
        \emph{Bottom:} The fit residuals.
        }
    \label{fig:photometry-rising_lightcurves}
\end{figure*}

\begin{table}[]
    \centering
    \caption{Best-fit parameters to the rising light curve of SN~2024pxl.}\label{tab:nir-fits}
    \begin{tabular}{cccc}
        \hline \\
        Quantity & Value &  Units & Description\\ 
        \hline \hline \\
            $t_0$ & $60511.80_{-0.18}^{+0.16}$     & MJD  & Time of first light    \vspace{0.50mm} \\
            $n_{o{_1}}$ & $1.32_{-0.05}^{+0.06}$   &      & $o$-band rise slope    \vspace{0.50mm} \\
            $n_{o{_2}}$ & $-0.04_{-0.01}^{+0.01}$  &      & $o$-band decline slope \vspace{0.50mm} \\
            $A_o$ & $1.62_{-0.02}^{+0.02}$         & mJy  & $o$-band constant      \vspace{0.50mm} \\
            $n_{g{_1}}$ & $1.50_{-0.07}^{+0.08}$   &      & $g$-band rise slope    \vspace{0.50mm} \\
            $n_{g{_2}}$ & $-0.06_{-0.01}^{+0.01}$  &      & $g$-band decline slope \vspace{0.50mm} \\
            $A_g$ & $1.21_{-0.02}^{+0.01}$         & mJy  & $g$-band constant      \vspace{0.50mm} \\
            $n_{r{_1}}$ & $1.30_{-0.05}^{+0.06}$   &      & $r$-band rise slope    \vspace{0.50mm} \\
            $n_{r{_2}}$ & $-0.04_{-0.01}^{+0.01}$  &      & $r$-band decline slope \vspace{0.50mm} \\
            $A_r$ & $1.67_{-0.02}^{+0.02}$         & mJy  & $r$-band constant      \vspace{0.50mm} \\
            \hline \\
    \end{tabular}
\end{table}

\subsection{Rising Light Curve Fits}

To determine the power-law slope in a given filter for SN~2024pxl (assuming the same time of first light in all bands), we use a power-law model defined as 
\begin{equation}
    f(t) = 
    \begin{cases} 
        0 & t < t_0 \\
        \Lambda\left(\frac{t-t_0}{1+z}\right)^\alpha& t \geq t_0. 
    \end{cases}\label{eq:1pl}
\end{equation}
\noindent to fit the ATLAS and ZTF forced photometry, where
\begin{equation}
    \alpha \equiv \alpha_1\left(1+\frac{\alpha_2(t-t_0)}{(1+z)}\right).
\end{equation}
\noindent The value of $\alpha$ in a one-component power law (i.e., $\alpha=a$) is sensitive to the proximity of peak light the fitting data extends \citep[e.g.,][]{Vallely19, Vallely21}, the $\alpha_1$ parameter from the two-component model is not as sensitive to what percentage of the peak flux that is included in the fit range \citep{Vallely21}. We use a flexible power law to avoid this issue, fitting through 5~days after $g$-band maximum light. We also define $\Lambda$ as
\begin{equation}
    \Lambda = A\times10^{-\alpha}.
\end{equation}
\noindent This scale factor breaks the degeneracy between the multiplicative scale factor and power-law slope \citep{Miller20b}. We use the \verb|emcee| fitting package \citep{emcee} as described in \citet{Hoogendam24_TDE} with the results shown in Figure \ref{fig:photometry-rising_lightcurves}. The fitted time of first light is MJD $60511.8\pm0.2$, $2.6\pm0.2$~days before discovery, $3.7(4.7)\pm0.2$~days before our first optical (NIR) spectrum, and $13.8\pm0.2$~days before $g$-band maximum.

SN~2024pxl is best described as a smoothly rising SN~Ia with no excess flux and is a ``single'' rising SN~Ia in the \citet{Hoogendam24} morphological scheme. The smooth rise also rules out significant excess flux from interaction with a companion or the circumstellar medium. The $g$- and $r$-band rise slopes for SN~2024pxl are slightly steeper ($1.72\pm0.07$ and $1.34\pm0.05$, respectively) than SN~2020udy ($1.38\pm0.11$, $1.29\pm0.07$, respectively; \citealp{Maguire23}). The primary power-law parameters $n_{g_1}$ and $n_{r_1}$ have been shown to be generally consistent with single-power-law fits \citep{Vallely21}. To better compare SNe~2020udy and 2024pxl, we fit using a similar single-power-law method as \citet{Maguire23}. We find $g$- and $r$-band exponents of $\sim$1.12 and $\sim$1.24, respectively. The $g$-band changes significantly, whereas the $r$-band does not; this is most likely an effect of inflexible power laws depending on the amount of rise that is fit and the sparser sampling in the $g$ band. In general, SN~2024pxl is similar to SN~2020udy and consistent with the recent sample study by \citet{Magee25}, which finds that SNe~Iax have shallower rising light curve slopes than normal SNe~Ia.

\subsection{Color Curves}\label{sec:colors}

\begin{figure}
    \centering
    \includegraphics[width=0.975\linewidth]{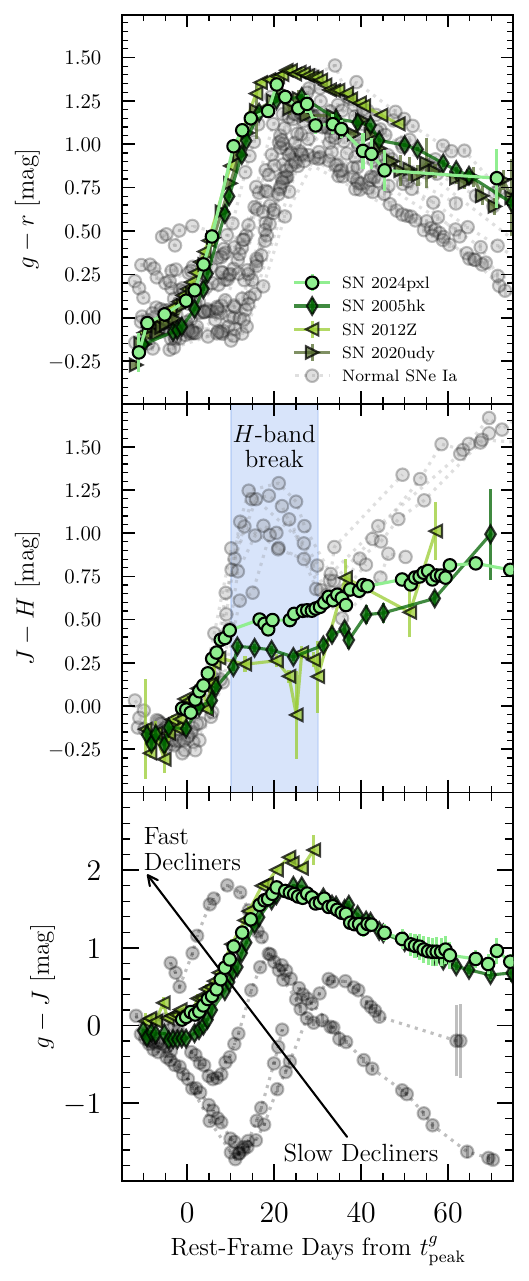} \\
    \caption{
        Color curves for SN~2024pxl compared against SNe 2005hk, 2012Z, and normal SNe~Ia from the Carnegie Supernova Project \citep{Stritzinger14, Yamanaka15, Krisciunas17}. The $g-r$ and $g-J$ color curves are corrected for Milky Way and host-galaxy extinction. The $J-H$ color curve is not corrected for extinction, as the effect is negligible \citep{Rieke85}. We estimate $E(g-J)\approx0.84$~mag assuming $A_J = 0.282\times A_V$ \citep{Rieke85} and $A_g = 1.2\times A_V$. SNe~2005hk, 2012Z, and 2020udy have minimal extinction \citep{Phillips07, Stritzinger15, Maguire23}.
        }
    \label{fig:photometry-colorcurve}
\end{figure}

Figure \ref{fig:photometry-colorcurve} shows the $g-r$ color curve of SN~2024pxl compared to three SNe~Iax (SNe~2005hk; \citealp{Phillips07}, 2012Z; \citealp{Stritzinger15}, and SN~2020udy; \citealp{Maguire23}) and normal SNe~Ia from the Carnegie Supernova Project \citep{Krisciunas17}. We highlight these three since they are well-observed, SNe~2005hk and 2012Z constitute our NIR colors comparison sample (see below), and SN~2020udy also has early-time $g-r$ colors. SN~2024pxl and the other SNe~Iax become redder even at early times. The general redward trend with no blueward evolution could be caused by SNe~Iax having a well-mixed ejecta with $^{56}$Ni out to the surface layers, rather than a centrally concentrated $^{56}$Ni distribution. Such an elemental distribution alters the energy distribution in the ejecta. This reduces the amount of stored energy in the central regions relative to normal SNe~Ia, and thus makes adiabatic cooling more dominant at early times. At the earliest epochs, SN~2024pxl and SN~2020udy have a ``knee'' in the color curve at $\sim$10~days before maximum light, becoming $\sim$0.20~mag redder in a day. SN~2024pxl matches the other two SNe~Iax near maximum light. 

To the best of our knowledge, we present the first analysis of the $J-H$ color curves of SNe~Iax. Only SNe~2005hk \citep{Phillips07} and 2012Z \citep{Yamanaka15} have sufficient NIR data for $J-H$ color curves to compare to SN~2024pxl. SN~2024pxl exhibits a similar evolution for the first two weeks after maximum light; however, during the $H$-band break region, it is redder than SN~2005hk and 2012Z. In the $g-J$ color curve, SNe~Iax start with $g-J\approx0$~mag, but they never become bluer, unlike normal SNe~Ia, which become significantly bluer as part of their evolution. Instead, SN~2024pxl and the other SNe~Iax become redder after maximum light in $g-J$, reaching their red maximum $\sim20$~days later. They slowly become bluer over time but remain several magnitudes redder than normal SNe~Ia. SN~2024pxl is $\sim$1~mag redder than both SNe~2005hk and 2012Z in $g-J$ before an extinction correction, but after, it has a similar color, supporting our extinction estimate.

Compared to normal SNe~Ia, SNe~Iax are on the red edge of the normal SNe~Ia $g-r$ colors throughout their evolution. SNe~Iax remain highly ionized for an extended period due to a combination of low ejecta velocities and $^{56}$Ni mixed throughout the ejecta. This is also the reason SNe~Iax lack the $r$-band secondary peak that is common in normal SNe~Ia and thought to be from recombination from \FeIII\ to \FeII\ \citep[e.g.,][]{Kasen06} The $J-H$ colors of normal SNe~Ia have a blue peak at $\sim$20~days after maximum light, corresponding to the phase of the $H$-band break in the NIR spectra \citep{Wheeler98, Ashall19a, Ashall19b}, but SN~2024pxl and the other SNe~Iax have a linear, blue $J-H$ evolution starting at $\sim$$+10$~days. The stronger $H$-band break in normal SNe~Ia may be because they generally produce more $^{56}$Ni than SNe~Iax, or that the velocity distribution of $^{56}$Ni differs between objects. By $\sim$$+30$~days, normal SNe~Ia and SN~2024pxl and the other SNe~Iax have similar $J-H$ colors again, but by $\sim$$+75$~days, normal SNe~Ia are redder than SN~2024pxl and the other SNe~Iax by nearly a magnitude. 

Finally, we also present a novel analysis of the $g-J$ colors for SNe~Iax compared to a representative sample of four SN~Ia from Figure 1 of \citet[][arranged by $\Delta m_{15}(B)$: SNe 2008hv, 2006ax, 2007on, and 2006mr]{Phillips12}. We limit our sample of normal SNe~Ia to these four to ensure the general trend within normal objects is visible and not obscured by too many objects. Normal SNe~Ia become bluer as they approach maximum light, with the fastest decliners reaching their bluest point near maximum light. 
There is no blueward $g-J$ evolution until $\sim$$+20$~days after maximum light. SNe~Iax may be a more homogeneous class with less intra-object variance than normal SNe~Ia. Our sample is small and may be biased, but SNe~2005hk, 2012Z, and 2024pxl span a range of luminosities and decline rates, suggesting that the sample is potentially representative. A larger sample of SNe~Ia and SNe~Iax with optical and NIR photometry (e.g., combining data from ground-based surveys and the upcoming Nancy Grace Roman Space Telescope \citealp{Spergel15}) would help improve this first-order explanation of the $g-J$ colors of SNe~Ia.

\section{Spectroscopic Analysis} \label{sec:spectroscopy}
We obtained fifteen optical spectra between $-10.1$ and $+18.8$ days and ten NIR spectra between $-9.1$ and $+45.7$~days relative to the $g$-band maximum. 

\subsection{Optical Spectra}\label{sec:optical_spectra}
\begin{figure}
    \centering
    \includegraphics[width=\linewidth]{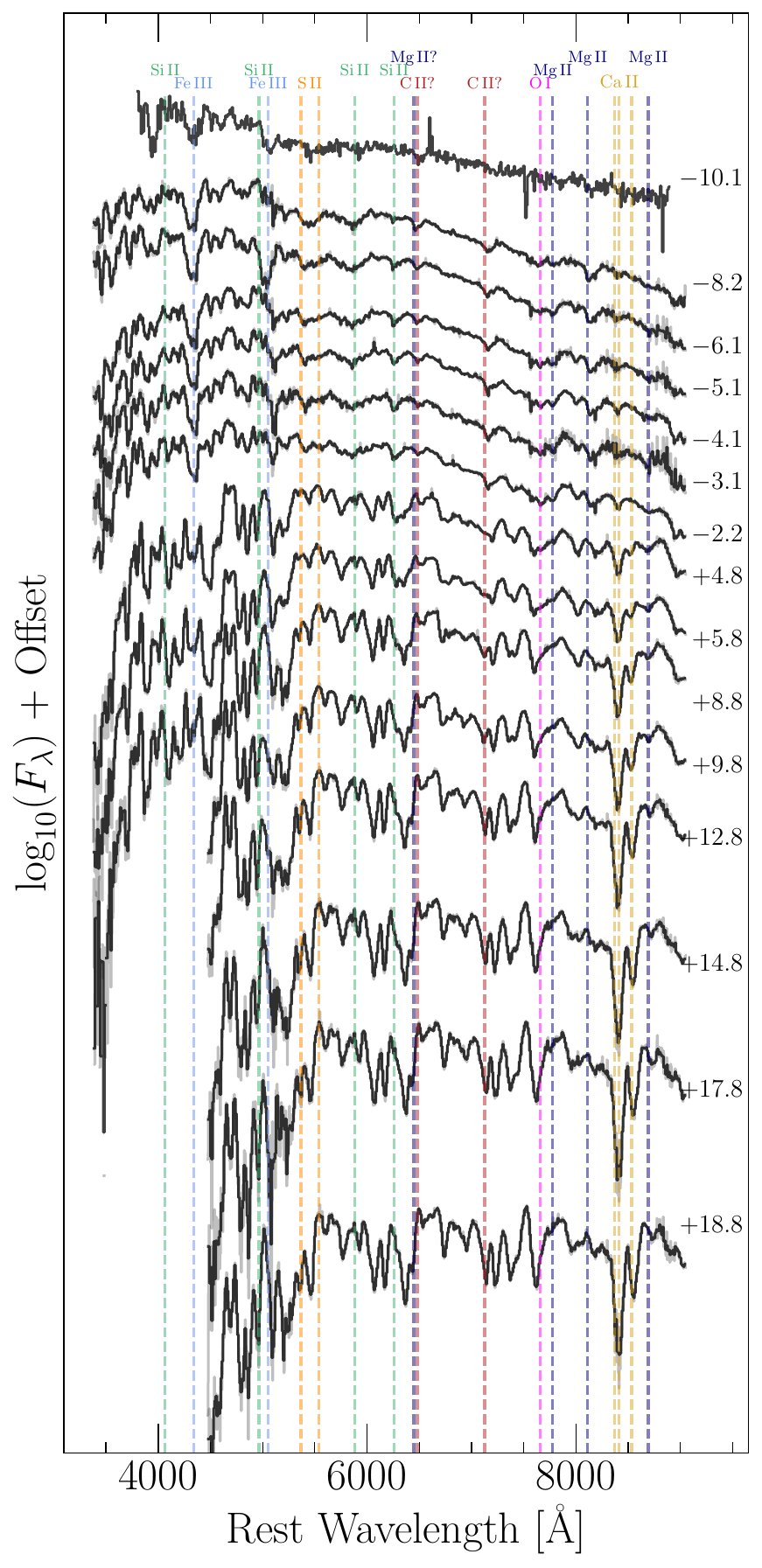} \\
    \caption{
        Optical spectroscopic time series for SN~2024pxl. Low SNR regions blue of 4500~\AA\ are not shown for the later spectra. Each spectrum is binned from the original data to 10~\AA\ wide bins. Phases are relative to the $g$-band maximum on MJD 60525.5. Dashed lines correspond to common SN~Iax features and are blueshifted by 4\,500~km~s$^{-1}$. 
        }
    \label{fig:spectra-optical}
\end{figure}

Figure \ref{fig:spectra-optical} shows the optical spectroscopic time series for SN~2024pxl, which has typical SN~Iax features at early times, with the strongest being \FeIII\ $\lambda\lambda$4404,~5129. \SII\ $\lambda\lambda$5454,~5640 and \SiII\ $\lambda$6355 emerge near maximum light along with a strong feature in the area of \CII\ $\lambda$6580, which is not uncommon for SNe~Iax \citep{Foley13} and often interpreted as C \citep[e.g.,][]{Maguire23}. 
There is a feature near \CII\ $\lambda$7234, similar to SN~2020udy with a tentative \CII~$\lambda7234$ feature identified \citep{Maguire23}. However, this feature has a higher velocity than \CII~$\lambda6580$. Alternatively, \citet{Szalai15} argue that this $\sim$7200~\AA\ feature is a blend of \FeII\ in SNe~Iax, which is more plausible given the velocity discrepancy with \CII~$\lambda6580$. 

\begin{figure}
    \centering
    \includegraphics[width=\linewidth]{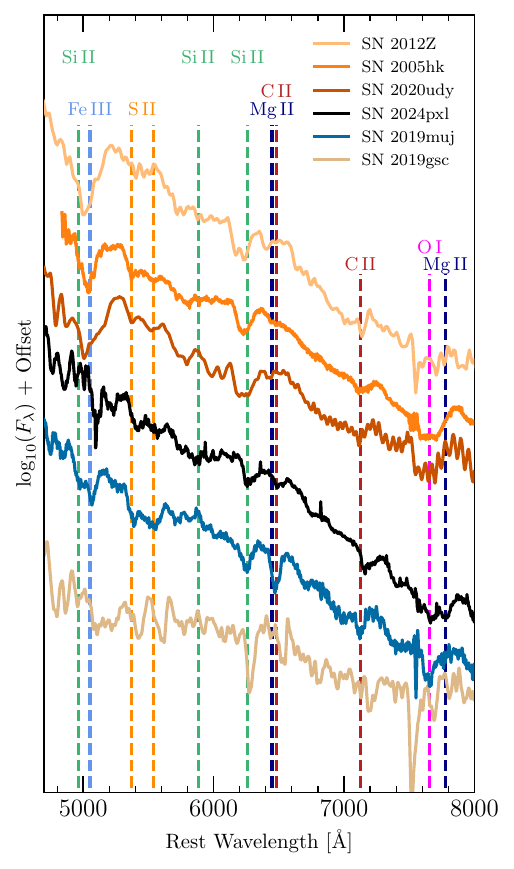} \\
    \caption{
        Comparison of the maximum light optical spectra of SN~2024pxl to other SNe~Iax. See the text for comparison object references. Dashed lines correspond to elements shown in Figure \ref{fig:spectra-optical} and are blueshifted by 4\,500~km~s$^{-1}$. 
        }
    \label{fig:spectra-optical_comp}
\end{figure}

Figure \ref{fig:spectra-optical_comp} compares the maximum light optical spectrum of SN~2024pxl with other SNe~Iax: SN~2005hk \citep{Phillips07}, SN~2012Z \citep{Stritzinger15}, SN~2019gsc \citep{Tomasella20}, SN~2019muj \citep{Barna21}, and SN~2020udy \citep{Maguire23}. The objects are arranged by absolute magnitude, with SN~2012Z being the most luminous and SN~2019gsc being the least luminous. The less luminous objects have weaker \FeIII~$\lambda5129$ and stronger \CII~$\lambda6580$, and the more luminous SNe~2005hk and 2020udy have stronger \FeIII~$\lambda5129$ and weaker \CII~$\lambda6580$. SN~2012Z does not follow this prescription exactly -- it has strong \FeIII\ and strong \CII~$\lambda6580$. This trend may suggest that even within the SNe~Iax class, there is a burning efficiency difference between more luminous objects having more burning (weaker \CII~$\lambda6580$) and lower luminosity objects having less burning (more \CII~$\lambda6580$). To confirm this, a systematic abundance tomography study of SNe~Iax is needed, which is beyond the scope of this work. 

\begin{figure}
    \centering
    \includegraphics[width=\linewidth]{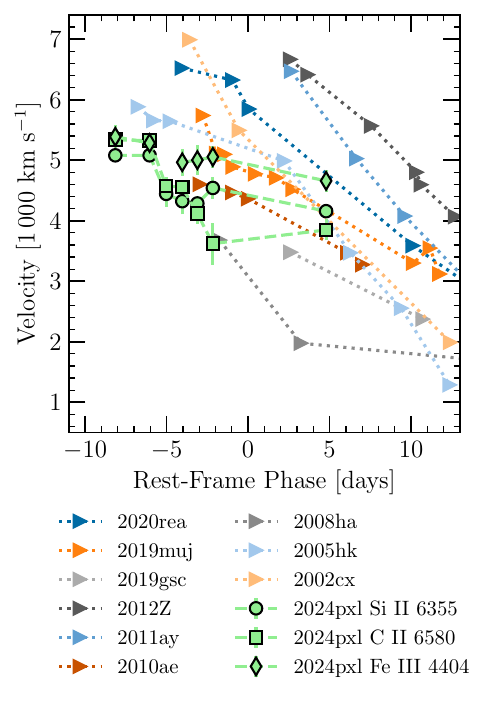} \\
    \caption{
        The velocity evolution of \SiII~$\lambda6355$ (light green circles), \CII~$\lambda6580$ (light green squares), and \FeIII~$\lambda4404$ (light green diamonds) for SN~2024pxl compared to the \SiII~$\lambda6355$ velocity evolution of other SNe~Iax (right-facing triangles).
        }
    \label{fig:spectra-velocities}
\end{figure}

Figure \ref{fig:spectra-velocities} shows the velocity evolution of \FeIII~$\lambda4404$, \SiII~$\lambda6355$, and \CII~$\lambda6580$ for SN~2024pxl compared to the \SiII~$\lambda6355$ velocities of other SNe~Iax. These three features trace different burning regions: C is a pure tracer of primordial unburnt material, Si traces C+O burning, and Fe traces Si burning into $^{56}$Ni (which decays into Fe). The similar velocity evolutions for the three features in SN~2024pxl suggest that the ejecta is efficiently mixed, containing unburnt, partially burnt, and fully burnt material in the same layers. This is a typical sign of deflagration burning. The first velocity point is highly uncertain due to the shallow line profiles, which may underestimate the velocities. The feature velocities for SN~2024pxl are lower than both normal SNe~Ia and bright SNe~Iax, such as SNe~2005hk and 2012Z \citep{Phillips07, Stritzinger15}, but similar to another SN~Iax: SN~2010ae \citep{Stritzinger14}.   

\subsection{NIR Spectra}\label{sec:NIR_Spectra}

\begin{figure*}
    \centering
    \includegraphics[width=\textwidth]{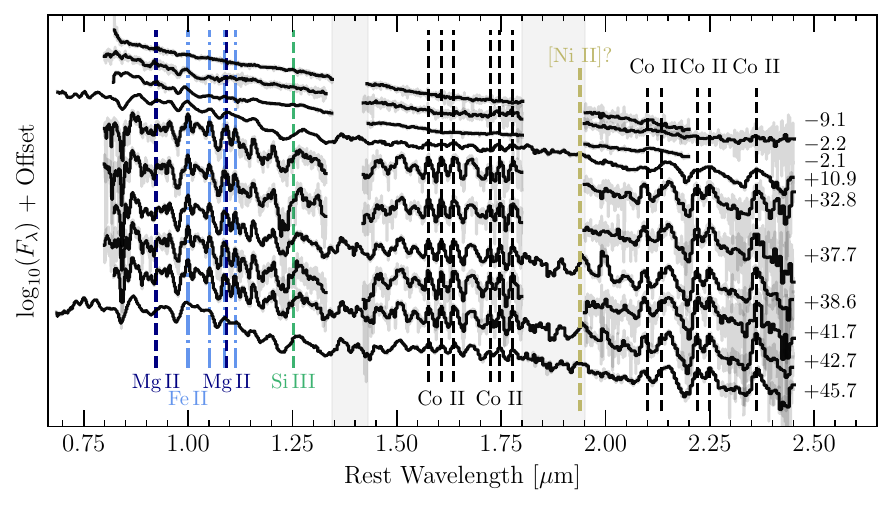} \\
    \caption{
        NIR spectroscopic time series for SN~2024pxl. Each spectrum (black) is binned from the original data (light grey) into bins 750~km~s$^{-1}$ wide. Spectral features are denoted with dashed lines, including \FeII, \MgII, \SiIII, and \CoII. The telluric regions are denoted with grey shading, and spectra with low S/N in this region have it masked. The phases are relative to maximum light. 
        }
    \label{fig:spectra-nir}
\end{figure*}

Figure \ref{fig:spectra-nir} shows our NIR spectral time series data. The early-time NIR spectra show a featureless continuum, similar to the other two SNe~Iax with spectra at a similar, but slightly later, phase \citep{Stritzinger15, Maguire23}. Near maximum light, features start to emerge, including intermediate mass elements (\MgII\ 0.9227,~1.0092,~1.0927~$\mu$m) and iron group elements (\FeII~0.9998, 1.0503, 1.0863, 1.1126~$\mu$m, and \CoII~1.5759,~1.6064,~and~1.6361~$\mu$m). These features increase in strength at $\sim$30~days after maximum light. Due to the low ejecta velocities, each feature is resolved rather than blended into a single feature as in normal SNe~Ia. 

The NIR regime covers strong features for two important elements: He and C. He is highly sensitive to the progenitor system, with He arising from an outer He shell or nondegenerate He-star companion. C is a pure tracer of unburnt material.
Our NIR spectra do not show conclusive He absorption features for both \HeI~1.083~$\mu$m and \HeI~2.058~$\mu$m. This is consistent with SNe~2005hk, 2012Z, 2015H, and 2020udy, which do not have He signatures \citep{Magee19, Maguire23} but stands in contrast to reported He detections in SNe 2004cs and 2007J \citep[][although it is unclear if these objects are bonafide SNe~Iax, see \citealp{White15, Foley16_Iax_Late_Time}]{Foley13}. 

The NIR \CI~1.0693~$\mu$m feature, the most common NIR C feature in SNe~Ia and one of the strongest, is also absent in our NIR spectra. Even well after maximum light, no \CI\ features are visible, similar to SN~2020udy \citep{Maguire23}. This suggests the ejecta remains hot enough to ionize C, producing \CII\ rather than \CI. The persistence of \CII~$\lambda6580$ through maximum light in our optical spectra indicates a well-mixed ejecta with unburnt, singly ionized C throughout.

Finally, we may see an emerging [\NiII]~1.939~$\mu$m feature, attributed to stable $^{58}$Ni since all the synthesized $^{58}$Ni has radioactively decayed by this phase \citep[e.g.,][]{Kumar25}. Previous work by \citet{Foley16_Iax_Late_Time} identified a [\NiII] feature at $\sim$200~days after the maximum light in the optical spectra of SNe~2008A and 2008ge. Generally, the presence of stable Ni is taken as an indicator of high-density burning \citep[e.g.,][]{Kwok23, DerKacy23a, DerKacy24, Ashall24}. Further nebular-phase observations will confirm if SN~2024pxl has [\NiII]~1.939~$\mu$m, similar to other SNe~Iax. 

\subsection{NIR Spectral Comparison of SNe Iax}
\begin{figure*}
    \centering
    \includegraphics[width=\linewidth]{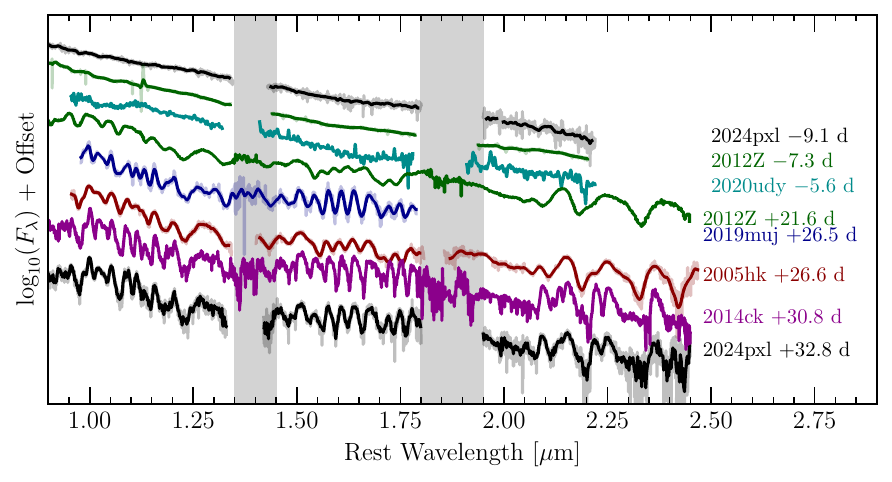} \\
    \caption{
        Comparison of the NIR spectra for SNe~Iax. Comparison objects are SN~2012Z \citep{Stritzinger15}, SN~2005hk \citep{Kromer13a}, SN~2014ck \citep{Tomasella16}, and SN~2019muj \citep{Barna21}. 
        }
    \label{fig:spectra-nir_comp}
\end{figure*}

NIR spectra of SNe~Iax, and in particular spectra over a week before maximum light, are rarer. We compare SN~2024pxl to other SNe~Iax with NIR spectra in Figure \ref{fig:spectra-nir_comp}. Our comparison objects are SN~2005hk \citep{Kromer13a}, SN~2012Z \citep{Stritzinger15}, SN~2019muj \citep{Barna21}, and SN~2020udy \citep{Maguire23}. 

At early times, roughly a week before maximum light, SN~2024pxl and the other SNe~Iax with early-time NIR spectra are similarly featureless. At maximum light, weak spectral features emerge, but the explosion remains hot, as the slow expansion velocities limit the efficacy of adiabatic expansion to cool the ejecta. It is not until nearly a month later that \CoII\ features dominate the $H$-(see \S\ref{sec:Hband_comp}) and $K$-bands. 

\subsection{The $H$-band Break in SNe Iax}\label{sec:Hband_comp}
The 20-30 day NIR Fe and Co emission in the $H$-band in SNe~Ia is denoted as the ``$H$-band break'' \citep{Wheeler98, Ashall19a, Ashall19b}. These features come from $^{56}$Ni decay products (Co and Fe) above the photosphere and are usually strongly blended in normal SNe~Ia \citep{Wheeler98}. However, in SNe~Iax, the low velocities minimize blending, allowing individual features to be resolved. These features trace the outer distribution of $^{56}$Ni, providing insights into the high-density burning and energetics of the explosion.

\begin{figure*}
    \centering
    \includegraphics[width=\textwidth]{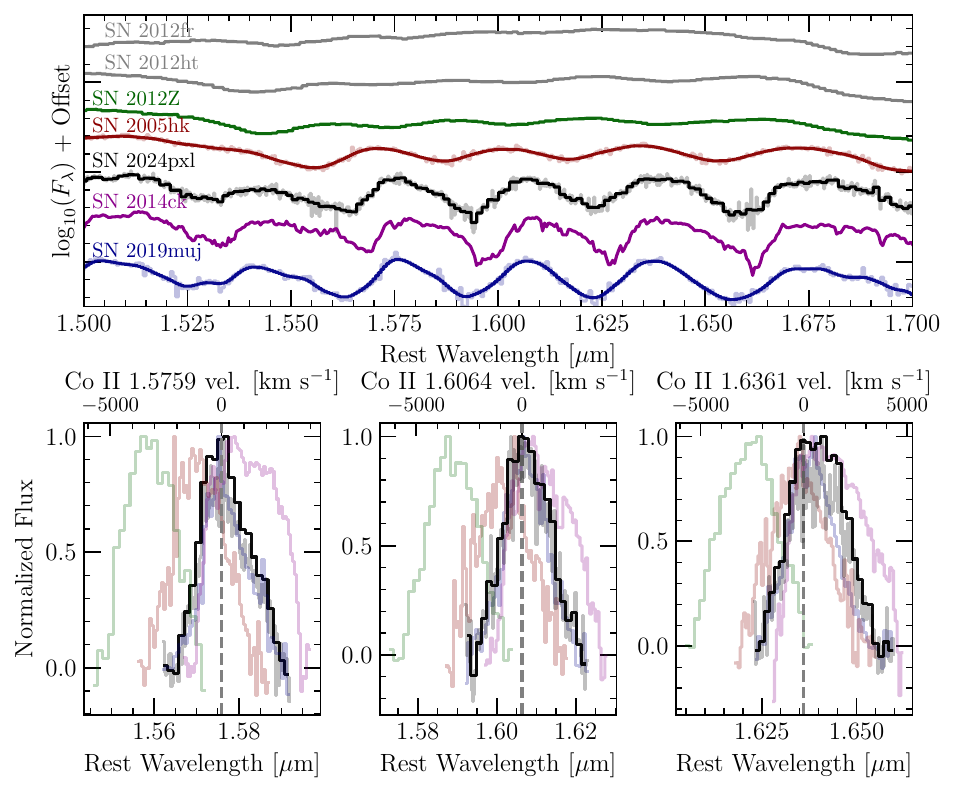} \\
    \caption{
        Comparison between the same SNe~Iax as in Figure \ref{fig:spectra-nir_comp}. Objects are arranged by absolute magnitude from top to bottom. The \CoII\ velocity correlates with absolute magnitude, with more luminous SNe~Iax having more blueshifted \CoII. We include two normal SNe~Ia from \citet{Lu23}: SNe 2012fr and 2012ht, to demonstrate blending in normal SNe~Ia. \emph{Top:} Spectra in the $H$-band region. \emph{Bottom:} Velocities for the three strongest \CoII\ features. The fluxes are normalized by the maximum flux value.
        }
    \label{fig:spectra-Hband_comp}
\end{figure*}

Figure \ref{fig:spectra-Hband_comp} shows the $\sim$30~day spectra for SNe~Iax and two normal SNe~Ia for comparison. The objects are arranged by peak magnitude. There is a correlation between peak magnitude and Doppler broadening of the NIR \CoII\ features in SNe~Iax, with more luminous SNe~Iax having higher velocities and generally broader line profiles \citep{Stritzinger14}. In addition to this known effect, we tentatively find a new one: the more luminous SNe~Iax have more blueshifted \CoII\ line profiles. It is not possible to detect this in the sample used by \citet{Stritzinger14} because their spectral sample is not large enough in the 20 to 30~day phase range. Comparing the blueshifting requires similar-epoch spectra such as those we compile and present in this work (ranging from $\sim$$+22$ to $\sim$$+33$~days).

 To obtain velocity estimates, we remove the pseudo-continuum using a linear fit of the edges of the emission feature and plot the pseudo-continuum subtracted emission features in the bottom panels of Figure \ref{fig:spectra-Hband_comp}. We do this for three \CoII\ lines: \CoII~1.5759,~1.6064,~and~1.6361~$\mu$m. SN~2012Z, a luminous SN~Iax, has a blueshift in the emission of $\sim$3\,000~km~s$^{-1}$, where as the slightly less luminous SN~2005hk has a shift only of $\sim$1\,000~km~s$^{-1}$. The least luminous SNe~Iax, 2014ck, 2019muj and 2024pxl, are near 0~km~s$^{-1}$. 

\begin{figure}
    \centering
    \includegraphics[width=\linewidth]{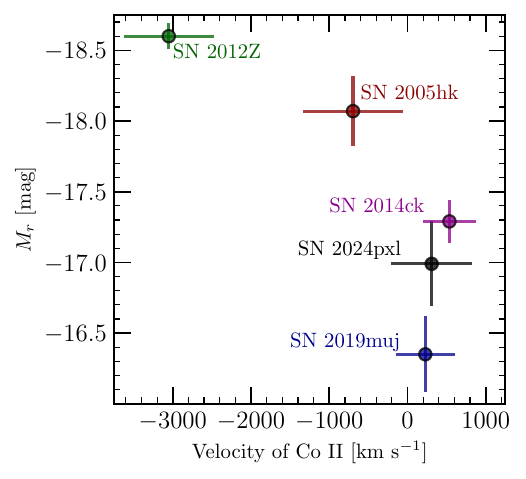} \\
    \caption{
        The $H$-band \CoII\ features the mean velocity shift against absolute $r$-band magnitude. The two most luminous SNe~Iax have higher $H$-band break \CoII\ velocities. The least luminous three are consistent with having the same velocity. 
        }
    \label{fig:spectra-Covel-vs-absmag}
\end{figure}

To demonstrate this effect better, we measure the peak Doppler velocity of the three \CoII\ peaks using a Gaussian fit to the pseudo-continuum-subtracted feature profile and calculate the median velocity. We adopt the standard deviation of the three peak velocity measurements as the uncertainty. Our results are shown in Figure \ref{fig:spectra-Covel-vs-absmag}. The more luminous SNe~Iax (SNe 2005hk and 2012Z) suggest there could be a correlation between velocity and absolute magnitude that saturates at lower luminosities (i.e., all SNe~Ia with $M_r\geq-17.5$~mag have velocities of $\sim$0~km~s$^{-1}$). The small sample size renders this correlation tentative; additional data are needed to improve the sample statistics and further analysis to clarify what physical origins may drive this trend. 

\subsection{0.3-2 micron SEDs}
\begin{figure}
    \centering
    \includegraphics[width=\linewidth]{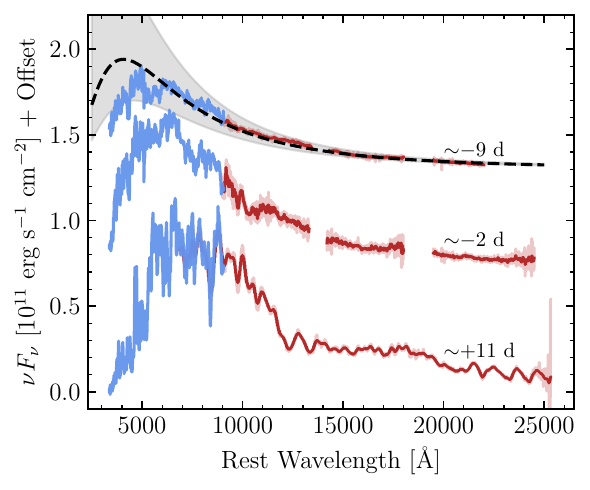} \\
    \caption{
        Optical (blue) and NIR (red) SED evolution for SN~2024pxl roughly one week before maximum light, at maximum light, and one week after maximum light. The dashed black line is the blackbody flux from a 9\,000 K blackbody. The shaded region upper and lower limits are 8\,000 and 10\,500 K, respectively.
        }
    \label{fig:spectra-sed}
\end{figure}

SN~2024pxl has three epochs with a complete optical to NIR SED at approximately one week before maximum light, at maximum light, and approximately one and a half weeks after maximum light. These are shown in Figure \ref{fig:spectra-sed}. The pre-peak SED shows a hot explosion dominated by the optical light. The NIR is featureless due to the high temperature of the explosion. At peak light, lines begin to form in the NIR as the explosion cools, but it is still hot and dominated by the optical. A week and a half later, the spectrum has cooled significantly, and the NIR contribution is more pronounced, with absorption features forming in the NIR as the temperature drops (although the explosion remains hot enough to ionize C). 

\section{Conclusion} \label{sec:conclusion}

We present multi-wavelength observations of the nearby SN~2024pxl: an intermediate luminosity SN~Iax with a typical light curve shape ($M_r = -16.99\pm0.32$~mag and $\Delta m_{15}(r)=0.69\pm0.05$~mag). Our extensive NIR photometry and spectroscopy make SN~2024pxl one of the most comprehensively observed SN~Iax in the NIR yet, alongside only SNe~2005hk and 2012Z. The rising light curve of SN~2024pxl does not show evidence for excess flux at early times, which is inconsistent with significant interaction with a companion star or circumstellar material.

SN~2024pxl has the earliest-ever NIR spectrum of a SN~Iax to date, at $-9.1$~days. The early-time NIR spectra are featureless, indicating a hot explosion at early times. At later times, $\sim$30 days after maximum light, the NIR spectra show resolved \CoII\ features in the $H$ and $K$ bands. Our optical spectra reveal a well-mixed, low-velocity, and highly ionized ejecta at early times characterized by \FeIII\ absorption features. SN~2024pxl exhibits a strong \CII~$\lambda6580$ feature throughout its photospheric phase. The smoothly rising light curves, well-mixed ejecta, and persistent C features are consistent with a deflagration explosion. 

Early-time and comprehensive NIR observations spanning the entire lifetimes of SNe~Iax offer additional tests of the leading deflagration models. The $J-H$ and $g-J$ color curves of SNe~Iax significantly differ from those of normal SNe~Ia and suggest that SNe~Iax begin adiabatic cooling immediately without centrally-concentrated $^{56}$Ni heating the ejecta. Around a month after maximum light, there may be a trend between the velocity of the \CoII\ features and the absolute magnitude of SNe~Iax, with more luminous objects showing a larger blueshift. 

As more SNe~Iax are discovered shortly after the explosion with multi-wavelength follow-up, larger population-level studies can be conducted at every phase, especially the earliest phases after the explosion.

\begin{acknowledgments}
We thank the anonymous referee for comments that improved the manuscript. We additionally thank Peter Hoeflich and Chris Burns for helpful discussions. 

% W.B.H. NSF 
This material is based upon work supported by the National Science Foundation Graduate Research Fellowship Program under Grant Nos. 1842402 and 2236415. Any opinions, findings, conclusions, or recommendations expressed in this material are those of the author(s) and do not necessarily reflect the views of the National Science Foundation.

%C.A. 
C.A. is supported by STScI grants (JWST-GO-02114, JWST-GO-02122, JWST-GO-04522, JWST-GO-03726, JWST-GO-6582, HST-AR-17555, JWST-GO-04217, JWST-GO-6023, JWST-GO-5290, JWST-GO-5057, JWST-GO-6677) and JPL-1717705.

% D.O.J.
D.O.J.\ acknowledges support from NSF grants AST-2407632 and AST-2429450, NASA grant 80NSSC24M0023, and HST/JWST grants HST-GO-17128.028, HST-GO-16269.012, and JWST-GO-05324.031, awarded by the Space Telescope Science Institute (STScI), which is operated by the Association of Universities for Research in Astronomy, Inc., for NASA, under contract NAS5-26555.

% B.J.S
The Shappee group at the University of Hawai'i is supported with funds from NSF (grants AST-1908952, AST-1911074, \& AST-1920392) and NASA (grants HST-GO-17087, 80NSSC24K0521, 80NSSC24K0490, 80NSSC24K0508, 80NSSC23K0058, \& 80NSSC23K1431).

% A.D. 
A.D.\ is supported by the European Research Council (ERC) under the European Union’s Horizon 2020 research and innovation programme (Grant Agreement No.\ 101002652).

% J. T. H. 
J.T.H. acknowledges support from NASA grant 80NSSC23K1431. 

% C. D. K
C.D.K. gratefully acknowledges support from the NSF through AST-2432037, the HST Guest Observer Program through HST-SNAP-17070 and HST-GO-17706, and from JWST Archival Research through JWST-AR-6241 and JWST-AR-5441. 

%% Gemini
Based on observations obtained at the international Gemini Observatory, a program of NSF NOIRLab, which is managed by the Association of Universities for Research in Astronomy (AURA) under a cooperative agreement with the U.S.\ National Science Foundation on behalf of the Gemini Observatory partnership: the U.S.\ National Science Foundation (United States), National Research Council (Canada), Agencia Nacional de Investigaci\'{o}n y Desarrollo (Chile), Ministerio de Ciencia, Tecnolog\'{i}a e Innovaci\'{o}n (Argentina), Minist\'{e}rio da Ci\^{e}ncia, Tecnologia, Inova\c{c}\~{o}es e Comunica\c{c}\~{o}es (Brazil), and Korea Astronomy and Space Science Institute (Republic of Korea). Data were taken by programs GN-2024A-Q-136, GN-2024A-Q-226, and GN-2024B-Q-109. This work was enabled by observations made from the Gemini North telescope, located within the Maunakea Science Reserve and adjacent to the summit of Maunakea. We are grateful for the privilege of observing the Universe from a place that is unique in both its astronomical quality and its cultural significance.

% IRTF
The Infrared Telescope Facility is operated by the University of Hawaii under contract 80HQTR24DA010 with the National Aeronautics and Space Administration.

% UKIRT 
UKIRT is owned by the University of Hawaii (UH) and operated by the UH Institute for Astronomy.

% Maunakea 
The authors wish to recognize and acknowledge the very significant cultural role and reverence that the summit of Maunakea has always had within the indigenous Hawaiian community.  We are most fortunate to have the opportunity to conduct observations from this mountain.

% ASAS-SN
We thank Las Cumbres Observatory and its staff for their continued support of ASAS-SN. ASAS-SN is funded in part by the Gordon and Betty Moore Foundation through grants GBMF5490 and GBMF10501 to the Ohio State University and also funded in part by the Alfred P. Sloan Foundation grant G-2021-14192.

% ATLAS
This work has used data from the Asteroid Terrestrial-impact Last Alert System (ATLAS) project. The Asteroid Terrestrial-impact Last Alert System (ATLAS) project is primarily funded to search for near-earth asteroids through NASA grants NN12AR55G, 80NSSC18K0284, and 80NSSC18K1575; byproducts of the NEO search include images and catalogs from the survey area. This work was partially funded by Kepler/K2 grant J1944/80NSSC19K0112 and HST GO-15889, and STFC grants ST/T000198/1 and ST/S006109/1. The ATLAS science products have been made possible through the contributions of the University of Hawaii Institute for Astronomy, Queen’s University Belfast, the Space Telescope Science Institute, the South African Astronomical Observatory, and the Millennium Institute of Astrophysics (MAS), Chile.
\end{acknowledgments}

% \begin{contribution}
%%This section gives authors the space to recognize author contributions. The text inside this environment is NOT counted towards the total word quanta. At a minimum, manuscripts are expected to include this text:

% All authors contributed equally to the Terra Mater collaboration.

%% But authors are expected to provide more specific details, e.g. 
%%
%%SC was responsible for writing and submitting the manuscript.
%%WWM came up with the initial research concept and edited the manuscript.
%%OTS obtained the funding and edited the manuscript.
%%EBF provided the formal analysis and validation. He also edited the manuscript.
%%GEH Supervised the undergraduates, wrote the software and administers the project github and Zenodo repositories.
%%
%% Authors can use the Contributor Role Taxonomy (CRediT) at
%% for ideas on how write a good statement tailored to their needs.

% \end{contribution}

\facilities{Swift(UVOT), IRTF(SpeX), Gemini:Gillett(GNIRS), UKIRT(WFCAM), UH:2.2m(SNIFS)}

% \software{astropy \citep{2013A&A...558A..33A,2018AJ....156..123A,2022ApJ...935..167A},  
%           Cloudy \citep{2013RMxAA..49..137F}, 
%           Source Extractor \citep{1996A&AS..117..393B}
%           }


% \appendix

%% For this sample we use BibTeX plus aasjournalv7.bst to generate the

\bibliography{sample7}{}
\bibliographystyle{aasjournalv7}



\end{document}